\documentclass[showpacs,amsmath,amssymb,twocolumn,pra,superscriptaddress,notitlepage]{revtex4-1}
\usepackage{graphicx}				
\usepackage{amssymb,amsmath}
\usepackage{braket}
\usepackage{subfigure}
\usepackage{hyperref} 
\usepackage{ulem}
\usepackage{color}

\begin{document}
\title{Proposed rapid detection of nuclear spins with entanglement-enhanced sensors}

\author{Hideaki Hakoshima 
}
\affiliation{Research Center for Emerging Computing Technologies, National institute of Advanced Industrial Science and Technology (AIST), Central2, 1-1-1 Umezono, Tsukuba, Ibaraki 305-8568, Japan}

\author{Yuichiro Matsuzaki 
}
\affiliation{Research Center for Emerging Computing Technologies, National institute of Advanced Industrial Science and Technology (AIST), Central2, 1-1-1 Umezono, Tsukuba, Ibaraki 305-8568, Japan}

\author{Toyofumi Ishikawa}
\email{toyo-ishikawa@aist.go.jp}
\affiliation{Research Center for Emerging Computing Technologies, National institute of Advanced Industrial Science and Technology (AIST), Central2, 1-1-1 Umezono, Tsukuba, Ibaraki 305-8568, Japan}

\begin{abstract}
Recently, there have been significant developments to detect nuclear spins with an nitrogen vacancy (NV) center in diamond. 
However, 
due to the nature of the short range dipole-dipole interaction,
it takes a long time to detect distant nuclear spins with the NV centers. 
Here, we propose a rapid detection of nuclear spins with an entanglement between the NV centers.
We show that
the necessary time to detect the nuclear spins with the entanglement is several orders of magnitude shorter than that with separable NV centers. 
Our result pave the way for new applications in nanoscale nuclear magnetic resonance spectroscopy.
\end{abstract}
\maketitle

\section{Introduction}


Nuclear magnetic resonance (NMR) spectroscopy is a widely applicable technique to characterize the behavior of atoms by measuring the magnetic fields generated from nuclear spins. However, a conventional NMR system with inductive coil sensors requires large sample volumes to obtain sufficient signals because of the weak magnetic fields and low thermal polarization of nuclear spins.

Qubit-based sensors, called quantum sensors, have attracted much attention for their ultra-high sensitivity beyond conventional magnetic sensors. Among the quantum sensors a nitrogen-vacancy (NV) quantum sensor, where the electron-spin state of a NV center in diamond is used as a spin qubit, has high spatial resolution in addition to high sensitivity of magnetic
field~\cite{RevModPhys.89.035002, Degen2008}.

With its advantages, an NV quantum sensor is expected to be a candidate for realizing NMR spectroscopy for small quantity of nuclear spins on micro- and nanometer- scales~\cite{Mamin557, Staudacher561, doi:10.1021/nl402286v, Lovchinsky503, Aslam67}.
To detect the nuclear spins with the NV centers, a thermal polarization of the nuclear spins are not required but a statistical polarization of the nuclear spins is used.
In this case, although the mean polarization of the nuclear spins is zero, there is a statistical fluctuation of the nuclear spins \cite{degen2009nanoscale,mamin2013nanoscale,staudacher2013nuclear}, and this induces measurable signals on the NV centers.

However, the NMR with NV centers
is useful only when the distance between the NV center and nuclear spins is around tens of nanometers or less.
The dipole-dipole interaction between the NV center and nuclear spins decreases by $r^{-3}$ where $r$ denotes the distance between them, and this limits the sensing distance of the NV centers.

On the other hand, entanglement enhanced sensing
has been attracted the attention of researchers due to the potentially high sensitivity.
To detect global magnetic fields, the sensing with $L$ probe qubits (or spins) provides with an estimation uncertainty to decrease by $L^{-1/2}$, which is called the standard quantum limit (SQL).
On the other hand, by using an entangled state with $L$ probe qubits, we can decrease the estimation uncertainty by $L^{-1}$ under ideal conditions. Even under the effect of decoherence, there are many protocols to use the entanglement enhanced sensing that surpasses  the SQL \cite{matsuzaki2011magnetic,chin2012quantum}.

Recently, a theoretical scheme to use the entanglement for the detection of a single spin was proposed
 \cite{hakoshima2020single,hakoshima2020efficient}.
 The crucial idea is to detect spatially inhomogeneous magnetic fields from the target single spin by using the entanglement enhanced magnetic field sensor. It was shown that, as long as the target spin is polarized during the protocol, the entanglement sensing  allows us to detect the target single spin much faster than the conventional sensing strategy with separable probe qubits
 \cite{hakoshima2020single,hakoshima2020efficient}.

 Here, we propose an entanglement enhanced NMR where we use an entanglement among the NV centers to detect a nuclear spin ensemble. Especially, we consider a case that the target nuclear spins are not polarized but prepared in a completely mixed state, which is the standard assumption 
 when we perform
 the NMR at room temperature.
 We show that a typical entanglement called a "GHZ state" among the NV centers
 can detect the nuclear spins by preforming a spin echo sequence on the NV centers.
 Moreover, when the distance between the NV centers and nuclear spins is around hundreds of nanometers, the necessary time to detect the nuclear spins with the entangled NV centers is 
 several orders of magnitude smaller than that with separable NV centers.
 Our results paves the way for many applications of the nano-scale NMR especially when the nuclear spins are located in a distant place.
 
\section{Setup}
\begin{figure}
\centering
\includegraphics[width=0.5\columnwidth,clip]{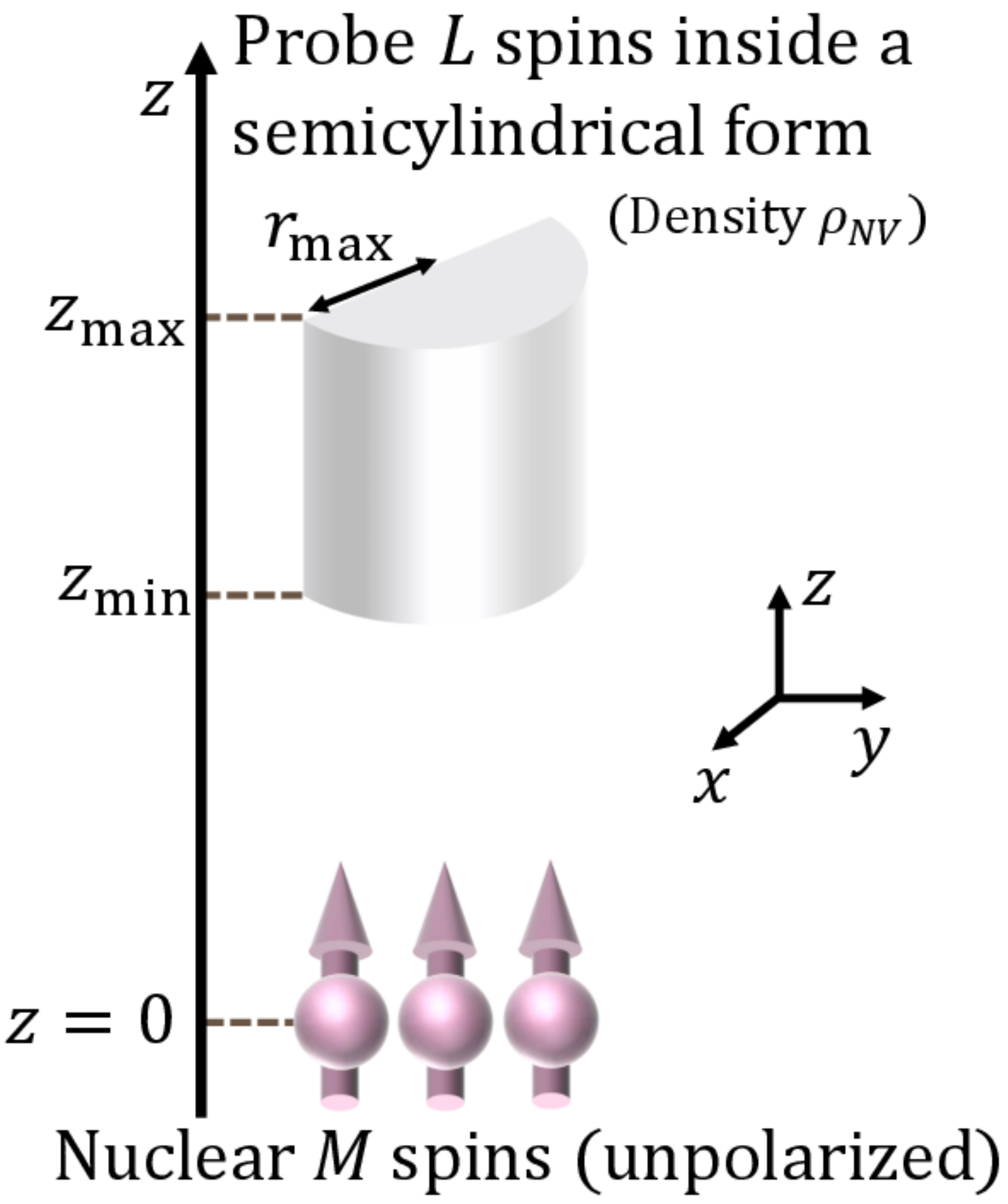}
\caption{Schematic of our scheme to detect unpolarized nuclear spins by using NV centers as a probe. The NV centers are uniformly distributed with a density $\rho_{\rm NV}$ inside a semicylindrical form ($\{ (x,y,z)|$ $x^2+y^2\le r_{{\rm max}}^2$, $z_{\rm min}\le z \le z_{\rm max}$, $x\ge 0\}$). We choose $z$-direction as a quantization axis for both the nuclear spins and the probe spins.
}
\label{fig:concept}
\end{figure}
As shown in Fig.~\ref{fig:concept}, we prepare $L$-qubit probe NV centers uniformly distributed with a density $\rho_{\rm NV}$ inside a semicylindrical form to detect  $M$-qubit nuclear spins.
Although the NV centers are spin-1 system, we can use only $|0\rangle $ and $|+1\rangle $ by ignoring $|-1\rangle $ by using a frequency selectivity. We set $\hbar=1$ throughout this paper.
The total Hamiltonian, which describes the Zeeman 
and the dipole-dipole interaction between the nuclear spins and the NV centers, is given by
\begin{align}
\hat{H}&=\hat{H}_{{\rm T}}+\hat{H}_{{\rm P}}+\hat{H}_{{\rm I}}\\
\hat{H}_{{\rm T}}&=\sum_{k=1}^M\frac{\omega^{({\rm T})} }{2}\hat{\sigma}_{z,k}^{({\rm T})}, \quad
\hat{H}_{{\rm P}}=\sum_{j=1}^L\frac{\omega^{({\rm P})}}{2}\hat{\sigma}_{z,j}^{({\rm P})},\\
\hat{H}_{{\rm I}}&=G\sum_{k=1}^M\sum_{j=1}^L\frac{\vec{\hat{\sigma}}^{({\rm T})}_k \cdot \vec{\hat{\sigma}}_{j}^{({\rm P})}-3 \frac{\vec{\hat{\sigma}}^{({\rm T})}_k \cdot\vec{r}_{kj}}{|\vec{r}_{kj}|} \frac{\vec{\hat{\sigma}}_{j}^{({\rm P})}\cdot\vec{r}_{kj}}{|\vec{r}_{kj}|}}{|\vec{r}_{kj}|^3},
\end{align}
where $\omega^{({\rm T})}$ ($\omega^{({\rm P})}$) is a resonant frequency of the nuclear  spins (NV centers), 
 $G=\frac{\mu_0 \gamma^{({\rm T})} \gamma^{({\rm P})} }{16\pi}$ 
 is 
 an effective coupling strength between the nuclear spins and NV centers, $\gamma^{({\rm T})}$ ( $\gamma^{({\rm P})}$) denotes magnetic moments of the nuclear spins (NV centers), 
$\vec{\hat{\sigma}}_{j}^{({\rm P})}=(\hat{\sigma}_{x,j}^{({\rm P})},\hat{\sigma}_{y,j}^{({\rm P})},\hat{\sigma}_{z,j}^{({\rm P})})$ is a 
set of the Pauli matrices of the NV centers at $\vec{r}_j=(x_j,y_j,z_j)$,
and
$\vec{r}_{kj}=\vec{r}_{k}-\vec{r}_{j}$ is a relative vector between a nuclear spin and a NV center.
In a rotating frame defined by $e^{i\hat{H}_{{\rm P}}t }$, we can use a rotating wave approximation and obtain
the following effective Hamiltonian 
\begin{align}
\hat{H}^{({\rm eff})}_{{\rm I}}=G\sum_{k,j}\left[A(\vec{r}_{kj})\hat{\sigma}_{x,k}^{({\rm T})}+B(\vec{r}_{kj})\hat{\sigma}_{y,k}^{({\rm T})}+C(\vec{r}_{kj})\hat{\sigma}_{z,k}^{({\rm T})}\right]\hat{\sigma}_{z,j}^{({\rm P})}
\end{align}
where we define $A(\vec{r}_{kj})=-\frac{3x_{kj}z_{kj}}{|\vec{r}_{kj}|^5}$, $B(\vec{r}_{kj})=-\frac{3y_{kj}z_{kj}}{|\vec{r}_{kj}|^5}$, and  $C(\vec{r}_{kj})=\frac{1}{|\vec{r}_{kj}|^3}\left(1-\frac{3z_{kj}^2}{|\vec{r}_{kj}|^2}\right)$.
Here, we assume $G\ll \omega^{({\rm T})}$ and we will calculate the leading order of $G/\omega^{({\rm T})}$. 

It is worth mentioning that the dynamics between the nuclear spins and NV centers was analyzed for the separable states by adopting a semi-classical model to treat the nuclear spins as a classical bath~\cite{PhysRevB.93.045425}. However, it is not clear whether such a semi-classical can be used when the NV centers to probe the nuclear spins are highly entangled. 
So we adopt a fully-quantized model to describe both nuclear spins and NV centers.

Since the most relevant decoherence of the NV center is dephasing, we consider the dephasing noise.
For a single-qubit state $\hat{\rho}$,
the dephasing 
map $\mathcal{E}^{(1)}$ 
is defined by \cite{de2010universal,PhysRevA.101.052303}
\begin{align}
\mathcal{E}^{(1)}[\hat{\rho}]=\frac{1+e^{-\left(\frac{t}{T_2}\right)^3}}{2}\hat{\rho}+ \frac{1-e^{-\left(\frac{t}{T_2}\right)^3}}{2}\hat{\sigma}_z \hat{\rho} \hat{\sigma}_z,
\label{eq:dephasingmap}
\end{align}
where $t$ is an interaction time and $T_2$ is a dephasing time.
For an 
$L$-qubit state, 
the dephasing map 
independently acts on each qubit:
$\mathcal{E}=\mathcal{E}^{(1)}\circ \mathcal{E}^{(1)}\circ \cdots \circ \mathcal{E}^{(1)}$. 

\section{Conventional protocol: separable state}
We describe how to detect the unpolarized nuclear spins with separable states through the periodic dynamical decoupling sequences.
The conventional protocol is shown as below:
In step 1, prepare the initial state $\frac{\mathbb{I}_{{\rm T}}}{2^M}\bigotimes_{j=1}^L \ket{+}_j\bra{+}_j$. Here, $\mathbb{I}_{{\rm T}}$ denotes a completely mixed state of the nuclear spins.
In step 2, let the state evolve for a time $\tau$ with the effective total Hamiltonian $\hat{H}_{{\rm T}}+\hat{H}^{({\rm eff})}_{{\rm I}}$ under the dephasing noise, and perform a $\pi$ pulse on all the probe spins.
In step 3, repeat step 2 $n_{\rm DD}$ times. For simplicity, we only consider that $n_{\rm DD}$ is a odd number. 
In step 4, measure the probe spins with the operator $\mathbb{I}_{{\rm T}}\otimes \hat{M}_x$, where $\hat{M}_x=\sum_{j=1}^L \hat{\sigma}_{x,j}^{({\rm P})}$.
In the final step, repeat 1-4 steps $N_m$ times.

The expectation value $\braket{\hat{M}_x}$ with dephasing noise
(see the derivation in Appendix A)
can be calculated as
\begin{align}
\braket{\hat{M}_x}=&e^{-2\left(\frac{t}{T_2^{\rm DD}}\right)^3}\left[ L-\frac{32G^2\sin^4{\frac{\omega^{({\rm T})}\tau}{2}}}{[\omega^{({\rm T})}]^2}\left[\frac{\sin{ \omega^{({\rm T})} t}}{\sin{\omega^{({\rm T})} \tau}}\right]^2\Gamma_{M,L}^{{\rm (sep)}}\right],
\label{eq:expectationvaluedynamical}
\end{align}
where $t=(n_{\rm DD}+1)\tau$ is an interaction time, $T_2^{\rm DD}=T_2^{\rm echo}\times (n_{\rm DD})^{2/3}$ 
\cite{klauder1962spectral,de2010universal,PhysRevB.85.155204}, and  
$
\Gamma_{M,L}^{{\rm (sep)}}=\sum_{k=1}^M\sum_{j=1}^L[A(\vec{r}_{kj})]^2+[B(\vec{r}_{kj})]^2
$ is a geometric factor, which is essentially the same result obtained in the previous study~\cite{PhysRevB.93.045425}.
We assume that a distance between nuclear spins is sufficiently small.
By assuming $z_{\rm min}\gg (\rho_{\rm NV})^{-1/3}$, we take a continuous limit
\begin{align}
\Gamma_{M,L}^{{\rm (sep)}}&\simeq M  \Gamma_{1,L}^{{\rm (sep)}}\simeq M\rho_{\rm NV} \int dV \frac{9(x^2+y^2)z^2}{(x^2+y^2+z^2)^5},
\label{eq:separablegeometric}
\end{align}
where $\rho_{\rm NV}$ is a 
density of the probe NV centers, as shown in Fig.~\ref{fig:concept}.
We can calculate the integral:
$
\int dV \frac{9(x^2+y^2)z^2}{(x^2+y^2+z^2)^5}
=\frac{3\pi }{8}z_{\rm min}^{-3} F_{1,L}^{{\rm (sep)}}(\tilde{r}_{{\rm max}},\tilde{z}_{{\rm max}})$,
although an analytical expression of 
is too complicated to be included here (the explicit form is in Appendix~\ref{sectionapp:Explicit}).
The variance is obtained as 
$
\braket{(\hat{M}_x)^2}-\braket{\hat{M}_x}^2=L(1-e^{-2\left(\frac{ t}{T_2^{\rm DD}}\right)^3}) 
$. We can calculate the signal-to-noise ratio from these results.
The signal for the case of separable states is given by
$
S_{\rm DD}=|\braket{\hat{M}_x}_{G=0}-\braket{\hat{M}_x}|
$, where $\braket{\hat{M}_x}_{G=0}=Le^{-\left(\frac{ t}{T_2^{\rm DD}}\right)^3}$.
The noise is given by
$
N_{\rm DD}=\sqrt{\frac{\braket{(\hat{M}_x)^2}-\braket{\hat{M}_x}^2}{N_m}}
$,
where $N_m=T/t$ is the number of measurement and $T$ is the total measurement time.
To detect the nuclear spins, $\frac{S_{\rm DD}}{N_{\rm DD}}\geq 1$ is required, and a minimum detectable time to satisfy $\frac{S_{\rm DD}}{N_{\rm DD}}=1$ is calculated as
$
T_d^{\rm (DD)}
= f^{\rm (DD)}(\tau) \times \frac{L}{[\Gamma_{M,L}^{{\rm (sep)}}]^2}
$,
where
$
f^{\rm (DD)}(\tau)=t\times \frac{e^{2\left(\frac{ t}{T_2^{\rm DD}}\right)^3}-1}{\sin^8{\frac{\omega^{({\rm T})}\tau}{2}}\left[\frac{\sin{ \omega^{({\rm T})} t}}{\sin{\omega^{({\rm T})} \tau}}\right]^4}\times \left(\frac{[\omega^{({\rm T})}]^2}{32G^2}\right)^2
$.
We need to optimize 
(i)$f^{\rm (DD)}(\tau)$ and (ii) $\frac{L}{[\Gamma_{M,L}^{{\rm (sep)}}]^2}$. 
We can minimize the first term by choosing optimal $\omega$ and $\tau$ so that 
a resonant condition of
$\tau= \pi/\omega^{({\rm T})}$ should be satisfied, and obtain $f^{\rm (DD)}(\tau)=\frac{\pi^4}{32(n_{\rm DD})^2(T_2^{\rm echo})^3G^4}$.
Also, for fixed $z_{\rm min}$ and $\rho_{\rm NV}$, we can numerically minimize the second term $
\frac{L}{[\Gamma_{1,L}^{{\rm (sep)}}]^2}
=\frac{32}{9\pi}\times \frac{z_{\rm min}^{9}}{\rho_{\rm NV} }\times\frac{\tilde{r}_{{\rm max}}^2[\tilde{z}_{{\rm max}}-1]}{\left[F_{1,L}^{{\rm (sep)}}(\tilde{r}_{{\rm max}},\tilde{z}_{{\rm max}})\right]^2}
$ by choosing $\tilde{r}_{{\rm max}}\simeq 1.16$ and  $\tilde{z}_{{\rm max}}\simeq 1.29$.
After these optimization, we obtain the optimized minimum detectable time
\begin{align}
T_d^{\rm (DD)}
&=c^{\rm (DD)}\times\frac{1}{(T_2^{\rm echo})^3G^4}\times \frac{z_{\rm min}^{9}}{\rho_{\rm NV}M^2} \frac{1}{(n_{\rm DD})^2},
\label{eq:minimumdetectabletimedd}
\end{align}
where $c^{\rm (DD)}\simeq 0.0536$.
\begin{figure}
\centering
\includegraphics[width=1.0\columnwidth,clip]{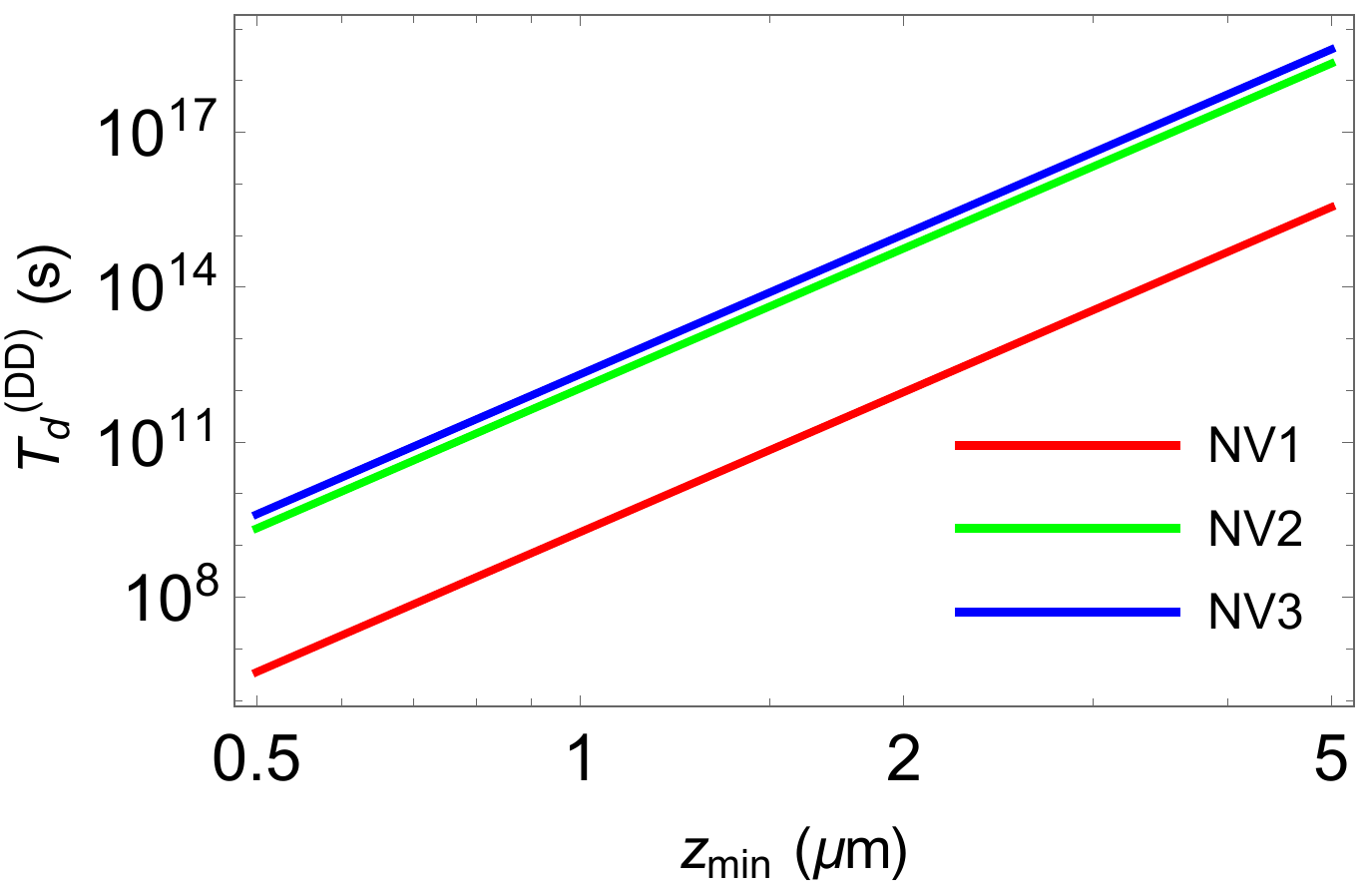}
\caption{Minimum detectable time $T_d^{\rm (DD)}$ with separable states
in Eq.~(\ref{eq:minimumdetectabletimedd}) against $z_{\rm min}$ with $n_{\rm DD}=63$. All the nuclear spins are protons ($\gamma^{({\rm T})}=2\pi\times 42$~MHz T$^{-1}$), and the number of the nuclear spins are $M=1.25\times 10^6$ (i.e. density $1.0\times 10^{22}$~cm$^{-3}$ inside a cube with a linear length $50$~nm). We employ the parameters of the NV centers realized in the previous experiment~\cite{doi:10.7566/JPSJ.89.054708}: (NV1) $T_2^{\rm echo}=8.3\times 10^{-5}$~s, $\rho_{\rm NV}=1.1\times10^{17}$~cm$^{-3}$; (NV2) $T_2^{\rm echo}=4.5\times 10^{-6}$~s, $\rho_{\rm NV}=1.1\times10^{18}$~cm$^{-3}$; (NV3) $T_2^{\rm echo}=3.1\times 10^{-4}$~s, $\rho_{\rm NV}=1.8\times10^{18}$~cm$^{-3}$.}
\label{fig:separableNV1NV2NV3}
\end{figure}
\begin{figure}
\centering
\includegraphics[width=1.0\columnwidth,clip]{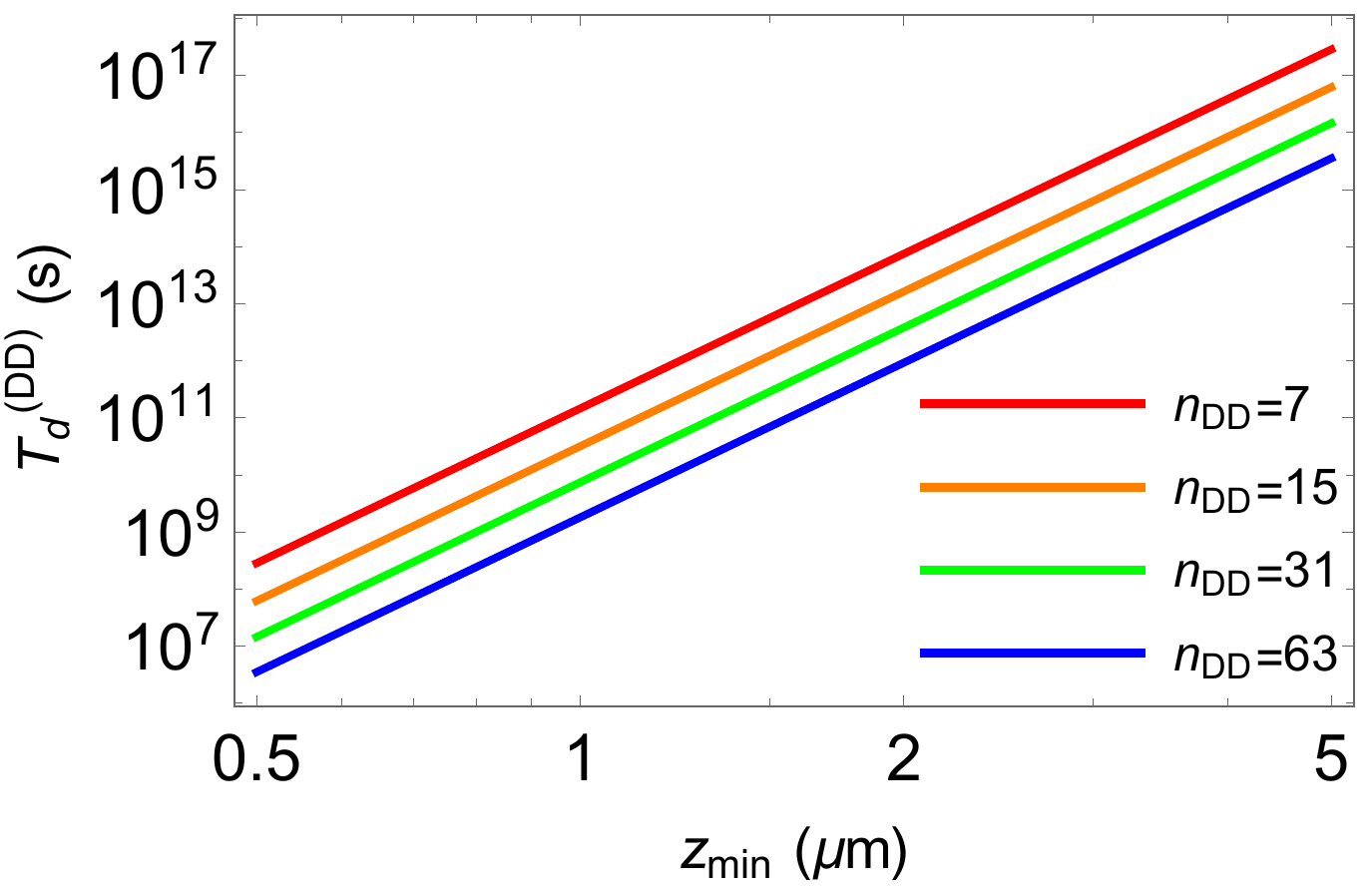}
\caption{Minimum detectable time $T_d^{\rm (DD)}$ with separable states
in Eq.~(\ref{eq:minimumdetectabletimedd}) against $z_{\rm min}$ for several $n_{\rm DD}$ with $T_2^{\rm echo}$ and $\rho_{\rm NV}$ fixed for NV3. All the other parameters are the same in Fig.~\ref{fig:separableNV1NV2NV3}.}
\label{fig:separablepulse}
\end{figure} 
In Figs.~\ref{fig:separableNV1NV2NV3} and \ref{fig:separablepulse}, we plot $T_d^{\rm (DD)}$ against $z_{\rm min}$. 
These results clearly show that, for $z_{\rm min}\geq 500$~nm, we cannot detect the
nuclear spins using the separable states in a realistic time (e.g., $10^9$s $\sim 30$ years).

\section{Our protocol: GHZ state}
%
Compared with the separable states, we show how to detect the unpolarized nuclear spins with GHZ states through the spin echo sequences, which corresponds to a special case of the dynamical decoupling at $n_{\rm DD}=1$.
Our protocol is shown as below:
In step 1, prepare the initial state $\frac{\mathbb{I}_{{\rm T}}}{2^M}\otimes \ket{{\rm GHZ}}\bra{{\rm GHZ}}$, where we define
\begin{align}
\ket{{\rm GHZ}}&= \frac{1}{\sqrt{2}}(\ket{00\cdots 0}+\ket{11\cdots 1}).
\end{align}
In step 2, let the state evolve for a time $\tau$ with the effective total Hamiltonian $\hat{H}_{{\rm T}}+\hat{H}^{({\rm eff})}_{{\rm I}}$ under dephasing noise, and perform a $\pi$ pulse on the probe spins.
In step 3, again, let the state evolve for a time $\tau$ with $\hat{H}_{{\rm T}}+\hat{H}^{({\rm eff})}_{{\rm I}}$, and perform a $\pi$ pulse on the probe spins.
In step 4, Measure the probe spins with the projection operator $\mathbb{I}_{{\rm T}}\otimes\ket{{\rm GHZ}}\bra{{\rm GHZ}}$. In the final step, repeat 1-4 steps $N_m$ times. 

We can calculate the measurement probability $p({\rm GHZ})$ (see the derivation in Appendix A)
in step 4:
\begin{align}
p({\rm GHZ})
&= \frac{1+e^{-L\left(\frac{2\tau}{T_2^{\rm echo}}\right)^3}}{2}-\frac{16G^2e^{-L\left(\frac{2\tau}{T_2^{\rm echo}}\right)^3}}{[\omega^{({\rm T})}]^2}\notag\\
&\times \sin^4{\frac{\omega^{({\rm T})}\tau}{2}}\times \Gamma_{M,L}^{{\rm (ent)}},
\label{eq:measurementprobabilityGHZ}
\end{align}
where 
$
\Gamma_{M,L}^{{\rm (ent)}}= \sum_{k=1}^M\sum_{j_1,j_2=1}^L[A(\vec{r}_{kj_{1}})A(\vec{r}_{kj_{2}})+B(\vec{r}_{kj_1})B(\vec{r}_{kj_2})]
$ is a geometric factor.
Under the same assumption of the separable states, $\Gamma_{M,L}^{{\rm (ent)}}$ can be approximated as
$
\Gamma_{M,L}^{{\rm (ent)}}\simeq M\times  \Gamma_{1,L}^{{\rm (ent)}}
$.
The geometric factor can be simplified as
\begin{align}
\Gamma_{1,L}^{{\rm (ent)}}&\simeq  (\rho_{\rm NV})^2F_{1,L}^{{\rm (ent)}}(\tilde{r}_{{\rm max}},\tilde{z}_{{\rm max}}),
\end{align}
although an analytical expression of 
is too complicated to be included here (the explicit form is in Appendix~\ref{sectionapp:Explicit}).
We can calculate the signal-to-noise ratio form the measurement probability $p({\rm GHZ})$.
The signal for the case of GHZ states is given by
$
S_{\rm GHZ}=|p({\rm GHZ})|_{G=0}-p({\rm GHZ})|, 
$
where $p({\rm GHZ})|_{G=0}=\frac{1}{2}+\frac{1}{2}e^{-L\left(\frac{2\tau }{T_2^{\rm echo}}\right)^2}$.
The noise is given by $N_{\rm GHZ}=\sqrt{\frac{p({\rm GHZ})[1-p({\rm GHZ})]}{N_m}}$.
Similarly, we define the minimum detectable time $T_d^{\rm (ent)}$ such that $\frac{S_{\rm DHZ}}{N_{\rm GHZ}}=1$ as
$
T_d^{{\rm (ent)}}
=f^{\rm (ent)}(\bar{\tau})\times \frac{L}{[\Gamma_{M,L}^{{\rm (ent)}}]^2},
$
where $\tau=\frac{\bar{\tau}}{L^{1/3}}$, $\omega^{({\rm T})}=L^{1/3} \bar{\omega}^{({\rm T})}$, and  
$
f^{\rm (ent)}(\bar{\tau})=\frac{2\bar{\tau}[e^{2\left(\frac{2\bar{\tau}}{T_2^{\rm echo}}\right)^3}-1]}{\sin^8{\frac{\bar{\omega}^{({\rm T})}\bar{\tau}}{2}}}\left(\frac{[\bar{\omega}^{({\rm T})}]^2}{32G^2}\right)^2=f^{\rm (DD)}(\bar{t})\left.\right|_{N_{\rm DD}=1}
$.
For fixed $z_{\rm min}$ and $\rho_{\rm NV}$, we can numerically minimize
$
\frac{L}{[\Gamma_{1,L}^{{\rm (ent)}}]^2}
=\frac{\pi}{2}\times \frac{z_{\rm min}^{3}}{\rho_{\rm NV}^{3} }\times\frac{[\tilde{r}_{{\rm max}}^2[\tilde{z}_{{\rm max}}-1]]}{\left[F_{1,L}^{{\rm (ent)}}(\tilde{r}_{{\rm max}},\tilde{z}_{{\rm max}})\right]^2}
$
by choosing $\tilde{r}_{{\rm max}}=5.05$ and  $\tilde{z}_{{\rm max}}=4.96$.
After these optimization, we obtain the optimized minimum detectable time
\begin{align}
T_d^{\rm (ent)}
&=c^{\rm (ent)}\times\frac{1}{(T_2^{\rm echo})^3G^4}\times \frac{z_{\rm min}^{3}}{\rho_{\rm NV}^{3} M^2},
\label{eq:minimumdetectabletimeent}
\end{align}
where $c^{\rm (ent)}=0.0107$. 
\begin{figure}
\centering
\includegraphics[width=1.0\columnwidth,clip]{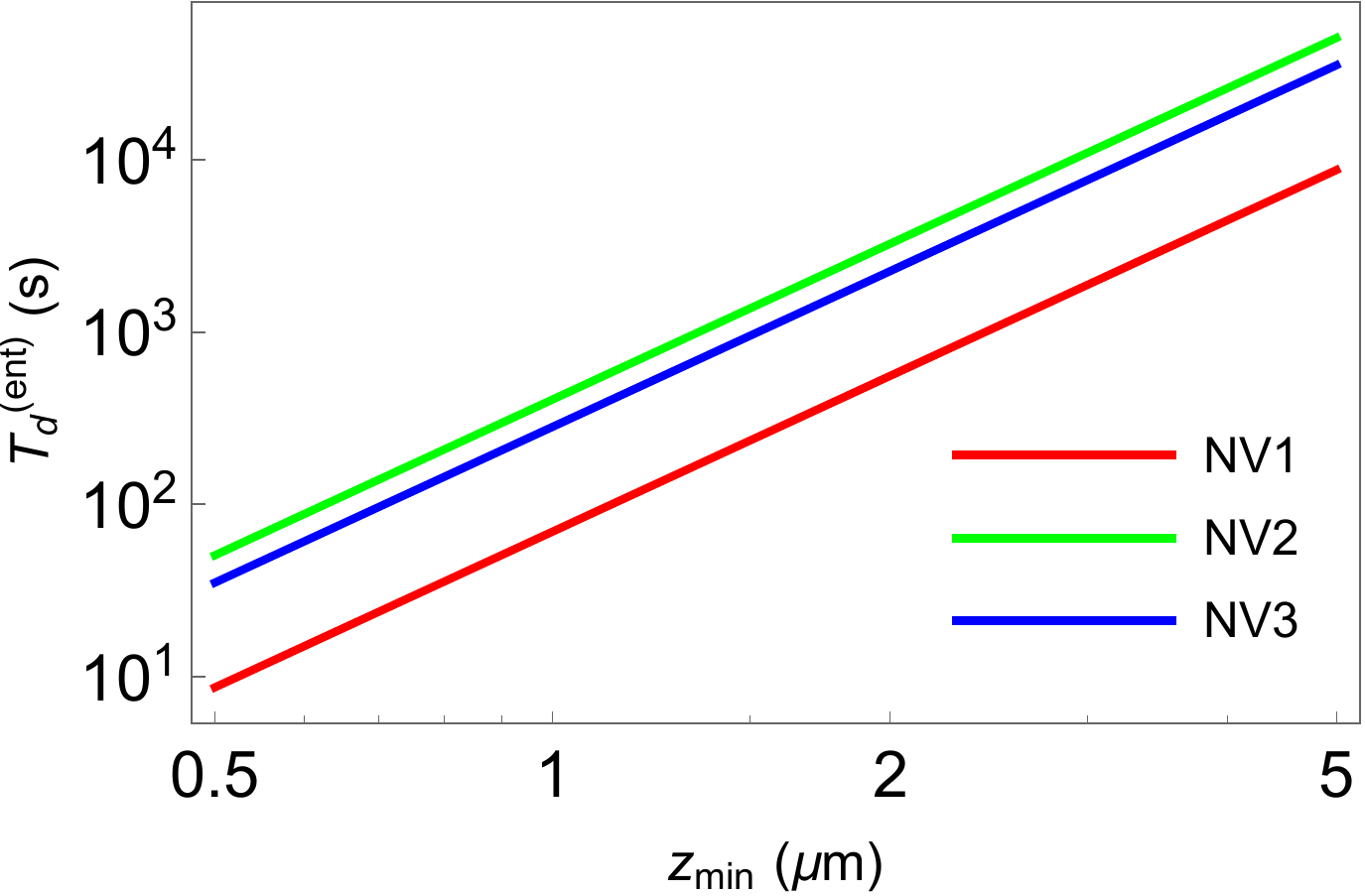}
\caption{Minimum detectable time $T_d^{\rm (ent)}$ with GHZ states
in Eq.~(\ref{eq:minimumdetectabletimeent}) against $z_{\rm min}$. All the parameters are the same in Fig.~\ref{fig:separableNV1NV2NV3}.
}
\label{fig:GHZNV1NV2NV3}
\end{figure}
In Fig.~\ref{fig:GHZNV1NV2NV3}, we plot $T_d^{\rm (ent)}$ against $z_{\rm min}$.
Compared with the separable states, we can efficiently detect the distant nuclear spins, as shown in Fig.~\ref{fig:GHZNV1NV2NV3}. Actually, we can confirm in Fig.~\ref{fig:GHZNV1NV2NV3} $T_d^{\rm (ent)}\sim 60$~s at $z_{\rm min}=1$~$\mathrm{\mu}$m for NV1, which is $10^7$ times faster than the case with the separable states.

\section{Conclusion}
In conclusion, we propose an entanglement enhanced detection of nuclear spins by using NV centers.
We
show that we can detect the nuclear spins with entangled NV centers
several orders of magnitude faster than the case with separable NV centers, especially when the distance between the nuclear spins and NV centers is around
or more than
hundreds of nanometers.
Our results suggest a novel approach to nanoscale NMR for detecting distant nuclear spins.

\begin{acknowledgments}
This work was supported by Leading Initiative for Excellent Young Researchers MEXT Japan and JST presto (Grant No. JPMJPR1919) Japan.
\end{acknowledgments}
\bibliographystyle{apsrev4-1}
\bibliography{NMRwithGHZ}

\appendix
\begin{widetext}

\section{Derivation for separable states and GHZ states}\label{app:derivation}
In this section, we will derive the expectation value in Eq.~(\ref{eq:expectationvaluedynamical}) and the measurement probability in Eq.~(\ref{eq:measurementprobabilityGHZ}).

\subsection{Second order perturbation formula}
We use the relation of the unitary operator up to the second order of $\varepsilon$:
\begin{align}
e^{-i(A-\varepsilon B)\tau}&=e^{-iA\tau}\left(1+i\varepsilon \int_0^\tau d\lambda e^{iA\lambda}Be^{-i(A-\varepsilon B)\lambda}\right)\\
&=e^{-iA\tau}\left(1+i\varepsilon \int_0^\tau d\lambda F(\lambda)-\varepsilon^2 \int_0^\tau d\lambda\int_0^\lambda d\xi  F(\lambda)F(\xi) \right)+O(\varepsilon^3),
\end{align}
where $F(\lambda)=e^{iA\lambda}Be^{-iA\lambda}$,
and
\begin{align}
e^{-i(A+\varepsilon B)\tau}&=\left(1-i\varepsilon \int_0^\tau d\lambda e^{-i(A+\varepsilon B)\lambda}Be^{iA\lambda}\right)e^{-iA\tau}\\
&=\left(1-i\varepsilon \int_0^\tau d\lambda F(-\lambda)-\varepsilon^2 \int_0^\tau d\lambda \int_0^\lambda d\xi F(-\xi)F(-\lambda) \right)e^{-iA\tau}+O(\varepsilon^3)
\end{align}
We obtain
\begin{align}
e^{-i(A-\varepsilon B)\tau}e^{-i(A+\varepsilon B)\tau}=&e^{-iA\tau}\left(1+i\varepsilon \int_0^{\tau}d\lambda [F(\lambda)-F(-\lambda)]\right.\notag\\
&\left.-\varepsilon^2 \int_0^{\tau}d\lambda \int_0^\lambda d\xi  [F(\lambda)F(\xi)+F(-\xi)F(-\lambda)]+\varepsilon^2 \int_0^{\tau}d\lambda \int_0^{\tau} d\xi  F(\lambda)F(-\xi)\right)e^{-iA\tau}+O(\varepsilon^3).\label{perturbationformula}
\end{align}
We will use this perturbation formula in the following.

\subsection{Derivation of the expectation value in Eq.~(\ref{eq:expectationvaluedynamical})}
Let us calculate the expectation value of $\hat{M}_x$ when the NV centers are prepared in separable states.
Without loss of generality (up to the second order perturbation), we calculate a case of a single-nuclear spin $M=1$.
We prepare the initial states with 
\begin{align}
\rho_{{\rm ini}}&=\frac{\mathbb{I}_{{\rm T}}}{2}\bigotimes_{j=1}^L \ket{+}_j\bra{+}_j\\
&=\frac{\ket{+}\bra{+}+\ket{-}\bra{-}}{2}\bigotimes_{j=1}^L \ket{+}_j\bra{+}_j\\
&=\frac{\ket{+}\bra{+}}{2}\bigotimes_{j=1}^L \ket{+}_j\bra{+}_j+\frac{\ket{-}\bra{-}}{2}\bigotimes_{j=1}^L \ket{+}_j\bra{+}_j.
\end{align}
So the initial state is a classical mixture of $\ket{+}\bra{+}\otimes_{j=1}^L \ket{+}_j\bra{+}_j $ and $\ket{-}\bra{-}\otimes_{j=1}^L \ket{+}_j\bra{+}_j $.
However, as shown below, 
the result with an initial state of $\ket{+}\bra{+}\otimes_{j=1}^L \ket{+}_j\bra{+}_j $ is the same as that with the other initial state of $\ket{-}\bra{-}\otimes_{j=1}^L \ket{+}_j\bra{+}_j $.
Therefore, we firstly consider the initial state $\ket{+}\bra{+}\otimes_{j=1}^L \ket{+}_j\bra{+}_j $, and secondly we show that the two initial states give the same result.

As we explain in the main text,
the conventional protocol is shown as below: In step 1, prepare the initial state $\ket{\psi(0)}=\ket{+}\bigotimes _{j=1}^L \ket{+}_j$.
In step 2, let the state evolve for a time $\tau$ with the effective total Hamiltonian $\hat{H}_{{\rm T}}+\hat{H}^{({\rm eff})}_{{\rm I}}$:\\
\begin{align}
\ket{\psi(\tau)}&=e^{-i(\hat{H}_{{\rm T}}+\hat{H}^{({\rm eff})}_{{\rm I}})\tau}\ket{+}\bigotimes _{j=1}^L \ket{+}_j\\
&=e^{-i(\frac{\omega^{({\rm T})} }{2}\hat{\sigma}_z^{({\rm T})}+\sum_{j=1}^L\hat{H}_j^{{\rm (T)}}\hat{\sigma}_{z,j}^{({\rm P})})\tau}\ket{+}\otimes\left(\frac{1}{2^{L/2}}\sum_s \ket{s}\right)\\
&=\frac{1}{2^{L/2}}\sum_s\left[e^{-i(\frac{\omega^{({\rm T})} }{2}\hat{\sigma}_z^{({\rm T})}+\sum_{j=1}^L\hat{H}_j^{{\rm (T)}}(-1)^{s_j})\tau}\ket{+}\right]\otimes \ket{s},
\end{align}
where we expand the states of the probe spins on the computational basis: $\ket{s}=\ket{s_1s_2\cdots s_L}$ ($s_i=0$ or $1$ or all $i$) and we define
\begin{align}
\hat{H}_j^{{\rm (T)}}=G\left[A(\vec{r}_{j})\hat{\sigma}_{x}^{({\rm T})}+B(\vec{r}_{j})\hat{\sigma}_{y}^{({\rm T})}+C(\vec{r}_{j})\hat{\sigma}_{z}^{({\rm T})}\right].
\end{align}
Moreover, perform a $\pi$ pulse on the probe spins:\\
\begin{align}
\ket{\psi'(\tau)}&=\left[\bigotimes_{j=1}^L\hat{\sigma}_{x,j}^{({\rm P})}\right]\ket{\psi(\tau)}\\
&=\frac{1}{2^{L/2}}\sum_s\left[e^{-i(\frac{\omega^{({\rm T})} }{2}\hat{\sigma}_z^{({\rm T})}+\sum_{j=1}^L\hat{H}_j^{{\rm (T)}}(-1)^{s_j})\tau}\ket{+}\right]\otimes \ket{s+1~{{\rm (mod~2)}}}.
\end{align}
Again, let the state evolve for a time $\tau$ with $\hat{H}_{{\rm T}}+\hat{H}^{({\rm eff})}_{{\rm I}}$, and perform a $\pi$ pulse on the probe spins:\\
\begin{align}
\ket{\psi'(2\tau)}&=\left[\bigotimes_{j=1}^L\hat{\sigma}_{x,j}^{({\rm P})}\right]e^{-i(\hat{H}_{{\rm T}}+\hat{H}^{({\rm eff})}_{{\rm I}})\tau}\ket{\psi'(\tau)}\\
&=\frac{1}{2^{L/2}}\sum_s\left[e^{-i(\frac{\omega^{({\rm T})} }{2}\hat{\sigma}_z^{({\rm T})}+\sum_{j=1}^L\hat{H}_j^{{\rm (T)}}(-1)^{s_j+1})\tau}e^{-i(\frac{\omega^{({\rm T})} }{2}\hat{\sigma}_z^{({\rm T})}+\sum_{j=1}^L\hat{H}_j^{{\rm (T)}}(-1)^{s_j})\tau}\ket{+}\right]\otimes \ket{s}\\
&=\frac{1}{2^{L/2}}\sum_s[\hat{M}_s \ket{+}]\otimes \ket{s}
\end{align}
where we define 
\begin{align}
\hat{M}_s&=e^{-i(\frac{\omega^{({\rm T})} }{2}\hat{\sigma}_z^{({\rm T})}-\sum_{j=1}^L\hat{H}_j^{{\rm (T)}}(-1)^{s_j})\tau}e^{-i(\frac{\omega^{({\rm T})} }{2}\hat{\sigma}_z^{({\rm T})}+\sum_{j=1}^L\hat{H}_j^{{\rm (T)}}(-1)^{s_j})\tau}\\
&=e^{-i\frac{\omega^{({\rm T})} \tau}{2}\hat{\sigma}_z^{({\rm T})}}(\mathbb{I}+\varepsilon \hat{M}_s^{(1)}+\varepsilon^2 \hat{M}_s^{(2)})e^{-i\frac{\omega^{({\rm T})} \tau}{2}\hat{\sigma}_z^{({\rm T})}}+O(\varepsilon^3).
\end{align}
Using the second order perturbation in Eq.~(\ref{perturbationformula}), $\hat{M}_s^{(1)}$ and $\hat{M}_s^{(2)}$ can be written as
\begin{align}
\varepsilon F_s(\lambda)=&\sum_{j=1}^L (-1)^{s_j}\vec{n}_j(\lambda)\cdot  \vec{\sigma}\\
\varepsilon\hat{M}_s^{(1)}=&i\varepsilon \int_0^{\tau}d\lambda [F_s(\lambda)-F_s(-\lambda)]\\
=&i\int_0^{\tau} d\lambda \sum_{j=1}^L(-1)^{s_j}[\vec{n}_j(\lambda) -\vec{n}_j(-\lambda)]\cdot  \vec{\sigma}\\
\varepsilon^2\hat{M}_s^{(2)}=&-\varepsilon^2 \int_0^{\tau}d\lambda \int_0^\lambda d\xi  [F_s(\lambda)F_s(\xi)+F_s(-\xi)F_s(-\lambda)]+\varepsilon^2 \int_0^{\tau}d\lambda \int_0^{t/2} d\xi  F_s(\lambda)F_s(-\xi)\\
=&\sum_{j_1,j_2=1}^L (-1)^{s_{j_1}+s_{j_2}}\int_0^{\tau}d\lambda \int_0^\lambda d\xi  (-1)\times [[\vec{n}_{j_1}(\lambda)\cdot  \vec{\sigma}][\vec{n}_{j_2}(\xi)\cdot  \vec{\sigma}]+[\vec{n}_{j_1}(-\xi)\cdot  \vec{\sigma}]\vec{n}_{j_2}(-\lambda)\cdot  \vec{\sigma}]\notag\\
&+\int_0^{\tau}d\lambda \int_0^{\tau} d\xi  [\vec{n}_{j_1}(\lambda)\cdot  \vec{\sigma}][\vec{n}_{j_2}(-\xi)\cdot  \vec{\sigma}],
\end{align}
where we define a vector
\begin{align}
\vec{n}_j(\lambda)&=G[(A(\vec{r}_j)\cos{(\omega^{({\rm T})}\lambda)}+B(\vec{r}_j)\sin{(\omega^{({\rm T})}\lambda)}),(-A(\vec{r}_j)\sin{(\omega^{({\rm T})}\lambda)}+B(\vec{r}_j)\cos{(\omega^{({\rm T})}\lambda)}),C(\vec{r}_j)].
\end{align}
In step 3, repeat the step 2 $N_{\rm DD}$ ($2N_{\rm DD}=n_{\rm DD}+1$) times:
\begin{align}
\ket{\psi'(N_{\rm DD}\tau)}=\frac{1}{2^{L/2}}\sum_s[(\hat{M}_s)^{N_{\rm DD}} \ket{+}]\otimes \ket{s}
\end{align}
Here, we calculate
\begin{align}
(\hat{M}_s)^{N_{\rm DD}}=&\left[e^{-i\frac{\omega^{({\rm T})} \tau}{2}\hat{\sigma}_z^{({\rm T})}}(\mathbb{I}+\varepsilon \hat{M}_s^{(1)}+\varepsilon^2 \hat{M}_s^{(2)})e^{-i\frac{\omega^{({\rm T})} \tau}{2}\hat{\sigma}_z^{({\rm T})}}\right]^{N_{\rm DD}}\\
=&e^{-i(N_{\rm DD}\omega^{({\rm T})} \tau)\hat{\sigma}_z^{({\rm T})}}+\varepsilon\sum_{l=1}^{N_{\rm DD}}e^{-i\frac{(2l-1)\omega^{({\rm T})} \tau}{2}\hat{\sigma}_z^{({\rm T})}}\hat{M}_s^{(1)}e^{-i\frac{(2(N_{\rm DD}-l)+1)\omega^{({\rm T})} \tau}{2}\hat{\sigma}_z^{({\rm T})}}\notag\\
&+\varepsilon^2\sum_{l_2> l_1}^{N_{\rm DD}}e^{-i\frac{(2l_1-1)\omega^{({\rm T})} \tau}{2}\hat{\sigma}_z^{({\rm T})}}\hat{M}_s^{(1)}e^{-i\frac{2(l_2-l_1)\omega^{({\rm T})} \tau}{2}\hat{\sigma}_z^{({\rm T})}}\hat{M}_s^{(1)}e^{-i\frac{(2(N_{\rm DD}-l_2)+1)\omega^{({\rm T})} \tau}{2}\hat{\sigma}_z^{({\rm T})}}\notag\\
&+\varepsilon^2\sum_{l=1}^{N_{\rm DD}}e^{-i\frac{(2l-1)\omega^{({\rm T})} \tau}{2}\hat{\sigma}_z^{({\rm T})}}\hat{M}_s^{(2)}e^{-i\frac{(2(N_{\rm DD}-l)+1)\omega^{({\rm T})} \tau}{2}\hat{\sigma}_z^{({\rm T})}}\\
=&e^{-i(N_{\rm DD}\omega^{({\rm T})} \tau)\hat{\sigma}_z^{({\rm T})}}+\varepsilon A_s^{(1)}+\varepsilon^2 A_s^{(2)},
\end{align}
where we define
\begin{align}
A_s^{(1)}=&\sum_{l=1}^{N_{\rm DD}}e^{-i\frac{(2l-1)\omega^{({\rm T})} \tau}{2}\hat{\sigma}_z^{({\rm T})}}\hat{M}_s^{(1)}e^{-i\frac{(2(N_{\rm DD}-l)+1)\omega^{({\rm T})} \tau}{2}\hat{\sigma}_z^{({\rm T})}}\\
A_s^{(2)}=&\sum_{l_2> l_1}^{N_{\rm DD}}e^{-i\frac{(2l_1-1)\omega^{({\rm T})} \tau}{2}\hat{\sigma}_z^{({\rm T})}}\hat{M}_s^{(1)}e^{-i\frac{2(l_2-l_1)\omega^{({\rm T})} \tau}{2}\hat{\sigma}_z^{({\rm T})}}\hat{M}_s^{(1)}e^{-i\frac{(2(N_{\rm DD}-l_2)+1)\omega^{({\rm T})} \tau}{2}\hat{\sigma}_z^{({\rm T})}}\notag\\
&+\sum_{l=1}^{N_{\rm DD}}e^{-i\frac{(2l-1)\omega^{({\rm T})} \tau}{2}\hat{\sigma}_z^{({\rm T})}}\hat{M}_s^{(2)}e^{-i\frac{(2(N_{\rm DD}-l)+1)\omega^{({\rm T})} \tau}{2}\hat{\sigma}_z^{({\rm T})}}.
\end{align}
In step 4, measure the expectation value as
\begin{align}
\braket{\hat{M}_x}&=\bra{\psi'(N_{\rm DD}\tau)}\mathbb{I}_{{\rm T}}\otimes \hat{M}_x\ket{\psi'(N_{\rm DD}\tau)}\\
&=\frac{1}{2^{L}}\sum_{s,s'}\bra{+}(\hat{M}_{s'}^\dagger)^{N_{\rm DD}} (\hat{M}_s)^{N_{\rm DD}} \ket{+}\bra{s'}\hat{M}_x\ket{s}
\end{align}
and we obtain
\begin{align}
(\hat{M}_{s'}^\dagger)^{N_{\rm DD}} (\hat{M}_s)^{N_{\rm DD}}
=& \mathbb{I}_{{\rm T}}+\varepsilon [e^{-i(N_{\rm DD}\omega^{({\rm T})} \tau)\hat{\sigma}_z^{({\rm T})}}(A_{s'}^{(1)})^\dagger + e^{i(N_{\rm DD}\omega^{({\rm T})} \tau)\hat{\sigma}_z^{({\rm T})}}A_s^{(1)}]\notag\\
&+\varepsilon^2 [(A_{s'}^{(1)})^\dagger A_s^{(1)}+(A_{s'}^{(2)})^\dagger e^{-i(N_{\rm DD}\omega^{({\rm T})} \tau)\hat{\sigma}_z^{({\rm T})}}+ e^{i(N_{\rm DD}\omega^{({\rm T})} \tau)\hat{\sigma}_z^{({\rm T})}}A_s^{(2)}]+O\left(\left(\frac{G}{[\omega^{({\rm T})}]}\right)^3\right).
\end{align}
Here, we use the following relation:
\begin{align}
\bra{s'}\hat{M}_x\ket{s}&=\sum_{j=1}^L\bra{s'}\sigma_{x,j}^{\rm (P)}\ket{s}=\sum_{j=1}^L \delta_{s_1's_1}\delta_{s_2's_2}\cdots \delta_{s_j's_j+1}\cdots \delta_{s_L's_L},
\end{align}
and it leads to these relations for $L>2$:
\begin{align}
\frac{1}{2^{L}}\sum_{s,s'}(-1)^{s_j}\bra{s'}\hat{M}_x\ket{s}&=\frac{1}{2^{L}}\sum_{s}(-1)^{s_j}=0\\
\frac{1}{2^{L}}\sum_{s,s'}(-1)^{s'_j}\bra{s'}\hat{M}_x\ket{s}&=\frac{1}{2^{L}}\sum_{s'}(-1)^{s'_j}=0\\
\frac{1}{2^{L}}\sum_{s,s'}(-1)^{s_{j_1}+s_{j_2}}\bra{s'}\hat{M}_x\ket{s}&=\frac{1}{2^{L}}\sum_{s,s'}(-1)^{s'_{j_1}+s'_{j_2}}\bra{s'}\hat{M}_x\ket{s}=L\delta_{j_1,j_2}\\
\frac{1}{2^{L}}\sum_{s,s'}(-1)^{s'_{j_1}+s_{j_2}}\bra{s'}\hat{M}_x\ket{s}&=\frac{1}{2^{L}}\sum_{s,s'}(-1)^{s_{j_1}+s'_{j_2}}\bra{s'}\hat{M}_x\ket{s}=(L-2)\delta_{j_1,j_2}.
\end{align}
From these relations, we can calculate the expectation value.
The first order of $O(\varepsilon)$ gives zero:
\begin{align}
&\frac{1}{2^{L}}\sum_{s,s'}\bra{+}\varepsilon[e^{-i(N_{\rm DD}\omega^{({\rm T})} \tau)\hat{\sigma}_z^{({\rm T})}}(A_{s'}^{(1)})^\dagger + e^{i(N_{\rm DD}\omega^{({\rm T})} \tau)\hat{\sigma}_z^{({\rm T})}}A_s^{(1)}] \ket{+}\bra{s'}\hat{M}_x\ket{s}\\
=&\frac{1}{2^{L}}\sum_{s,s'}\sum_{l=1}^{N_{\rm DD}}\bra{+} e^{i(N_{\rm DD}\omega^{({\rm T})} \tau)\hat{\sigma}_z^{({\rm T})}}e^{-i\frac{(2l-1)\omega^{({\rm T})} \tau}{2}\hat{\sigma}_z^{({\rm T})}}\varepsilon\hat{M}_s^{(1)}e^{-i\frac{(2(N_{\rm DD}-l)+1)\omega^{({\rm T})} \tau}{2}\hat{\sigma}_z^{({\rm T})}} \ket{+}\bra{s'}\hat{M}_x\ket{s}+c.c.\\
=&\frac{i}{2^{L}}\int_0^{\tau} d\lambda\sum_{j=1}^L\sum_{s,s'}(-1)^{s_j}\bra{s'}\hat{M}_x\ket{s}\sum_{l=1}^{N_{\rm DD}}\bra{+} e^{i\frac{(2(N_{\rm DD}-l)+1)\omega^{({\rm T})} \tau}{2}\hat{\sigma}_z^{({\rm T})}}  [\vec{n}_j(\lambda) -\vec{n}_j(-\lambda)]\cdot  \vec{\sigma} e^{-i\frac{(2(N_{\rm DD}-l)+1)\omega^{({\rm T})} \tau}{2}\hat{\sigma}_z^{({\rm T})}} \ket{+}+c.c.\\
=&0.
\end{align}
The second order of $O(\varepsilon^2)$ gives a non-zero contribution:
\begin{align}
&\frac{1}{2^{L}}\sum_{s,s'}\bra{+}\varepsilon^2 [(A_{s'}^{(1)})^\dagger A_s^{(1)}+(A_{s'}^{(2)})^\dagger e^{-i(N_{\rm DD}\omega^{({\rm T})} \tau)\hat{\sigma}_z^{({\rm T})}}+ e^{i(N_{\rm DD}\omega^{({\rm T})} \tau)\hat{\sigma}_z^{({\rm T})}}A_s^{(2)}]\ket{+}\bra{s'}\hat{M}_x\ket{s}
\end{align}
The first term gives
\begin{align}
&\frac{1}{2^{L}}\sum_{s,s'}\bra{+}\varepsilon^2 [(A_{s'}^{(1)})^\dagger A_s^{(1)}]\ket{+}\bra{s'}\hat{M}_x\ket{s}\\
=&\frac{1}{2^{L}}\sum_{s,s'}\bra{s'}\hat{M}_x\ket{s}\sum_{l_1,l_2}\bra{+}e^{i\frac{(2(N_{\rm DD}-l_1)+1)\omega^{({\rm T})} \tau}{2}\hat{\sigma}_z^{({\rm T})}} \left(\int_0^{\tau} d\lambda \sum_{j=1}^L(-1)^{s'_j}[\vec{n}_j(\lambda) -\vec{n}_j(-\lambda)]\cdot  \vec{\sigma}\right)^\dagger\notag\\
& e^{i\frac{(2l_1-2l_2)\omega^{({\rm T})} \tau}{2}\hat{\sigma}_z^{({\rm T})}}\left(\int_0^{\tau} d\lambda \sum_{j=1}^L(-1)^{s_j}[\vec{n}_j(\lambda) -\vec{n}_j(-\lambda)]\cdot  \vec{\sigma}\right)e^{-i\frac{(2(N_{\rm DD}-l_2)+1)\omega^{({\rm T})} \tau}{2}\hat{\sigma}_z^{({\rm T})}}\ket{+}\\
=&(L-2)\sum_{l_1,l_2}\sum_{j=1}^L\bra{+}e^{i\frac{(2(N_{\rm DD}-l_1)+1)\omega^{({\rm T})} \tau}{2}\hat{\sigma}_z^{({\rm T})}} \left(\frac{2G}{\omega^{({\rm T})}}[1-\cos{(\omega^{({\rm T})}\tau)}]\times [B(\vec{r}_j),-A(\vec{r}_j),0]\cdot  \vec{\sigma}\right)\notag\\
& e^{i\frac{(2l_1-2l_2)\omega^{({\rm T})} \tau}{2}\hat{\sigma}_z^{({\rm T})}}\left(\frac{2G}{\omega^{({\rm T})}}[1-\cos{(\omega^{({\rm T})}\tau)}]\times [B(\vec{r}_j),-A(\vec{r}_j),0]\cdot  \vec{\sigma}\right)e^{-i\frac{(2(N_{\rm DD}-l_2)+1)\omega^{({\rm T})} \tau}{2}\hat{\sigma}_z^{({\rm T})}}\ket{+}\\
=&(L-2)\frac{4G^2}{[\omega^{({\rm T})}]^2}[1-\cos{(\omega^{({\rm T})}\tau)}]^2\sum_{l_1,l_2}\sum_{j=1}^L\bra{+}e^{i\frac{(2(N_{\rm DD}-l_1)+1)\omega^{({\rm T})} \tau}{2}\hat{\sigma}_z^{({\rm T})}} \left( B(\vec{r}_j)\hat{\sigma}_x^{({\rm T})}-A(\vec{r}_j)\hat{\sigma}_y^{({\rm T})}\right)\notag\\
& e^{-i\frac{(2(N_{\rm DD}-l_1)+1)\omega^{({\rm T})} \tau}{2}\hat{\sigma}_z^{({\rm T})}} e^{i\frac{(2(N_{\rm DD}-l_2)+1)\omega^{({\rm T})} \tau}{2}\hat{\sigma}_z^{({\rm T})}} \left(B(\vec{r}_j)\hat{\sigma}_x^{({\rm T})}-A(\vec{r}_j)\hat{\sigma}_y^{({\rm T})}\right)e^{-i\frac{(2(N_{\rm DD}-l_2)+1)\omega^{({\rm T})} \tau}{2}\hat{\sigma}_z^{({\rm T})}}\ket{+}\\
=&(L-2)\frac{4G^2}{[\omega^{({\rm T})}]^2}[1-\cos{(\omega^{({\rm T})}\tau)}]^2\sum_{l_1,l_2}\sum_{j=1}^L\bra{+}(\vec{n}_{l_1}\cdot  \vec{\sigma})(\vec{n}_{l_2}\cdot  \vec{\sigma})\ket{+}\\
=&(L-2)\frac{4G^2}{[\omega^{({\rm T})}]^2}[1-\cos{(\omega^{({\rm T})}\tau)}]^2\sum_{l_1,l_2}\sum_{j=1}^L[\vec{n}_{l_1}\cdot\vec{n}_{l_2}+i \bra{+}(\vec{n}_{l_1}\times \vec{n}_{l_2})\cdot  \vec{\sigma}\ket{+}\\
=&(L-2)\frac{4G^2}{[\omega^{({\rm T})}]^2}[1-\cos{(\omega^{({\rm T})}\tau)}]^2\sum_{l_1,l_2}\sum_{j=1}^L[\vec{n}_{l_1}\cdot\vec{n}_{l_2}]\\
=&(L-2)\frac{4G^2}{[\omega^{({\rm T})}]^2}[1-\cos{(\omega^{({\rm T})}\tau)}]^2\sum_{l_1,l_2}\sum_{j=1}^L[(A(\vec{r}_j))^2+(B(\vec{r}_j))^2]\cos{[2(l_1-l_2)\omega^{({\rm T})} \tau]}\\
=&(L-2)\frac{16G^2}{[\omega^{({\rm T})}]^2}[\sin{\frac{\omega^{({\rm T})}\tau}{2}}]^4\left[\frac{\sin{(N_{\rm DD}\omega^{({\rm T})}\tau)}}{\sin{(\omega^{({\rm T})}\tau)}}\right]^2\Gamma_{1,L}^{{\rm (sep)}},
\end{align}
where we define a vector
\begin{align}
\vec{n}_{l}=& [B(\vec{r}_j)\cos{[(2(N_{\rm DD}-l)+1)\omega^{({\rm T})} \tau]}-A(\vec{r}_j)\sin{[(2(N_{\rm DD}-l)+1)\omega^{({\rm T})} \tau]}\notag\\
&,-[A(\vec{r}_j)\cos{[(2(N_{\rm DD}-l)+1)\omega^{({\rm T})} \tau]}+B(\vec{r}_j)\sin{[(2(N_{\rm DD}-l)+1)\omega^{({\rm T})} \tau]}],0]
\end{align}
and use the relation $\sum_{l_1,l_2}(\vec{n}_{l_1}\times \vec{n}_{l_2})=0$,
and we define a geometric factor
\begin{align}
\Gamma_{1,L}^{{\rm (sep)}}=\sum_{j=1}^L[(A(\vec{r}_j))^2+(B(\vec{r}_j))^2]
.
\end{align}
Moreover, the other terms give
\begin{align}
&\frac{1}{2^{L}}\sum_{s,s'}\bra{+}\varepsilon^2 [(A_{s'}^{(2)})^\dagger e^{-i(N_{\rm DD}\omega^{({\rm T})} \tau)\hat{\sigma}_z^{({\rm T})}}+ e^{i(N_{\rm DD}\omega^{({\rm T})} \tau)\hat{\sigma}_z^{({\rm T})}}A_s^{(2)}]\ket{+}\bra{s'}\hat{M}_x\ket{s}\\
=&\frac{\varepsilon^2}{2^{L}}\sum_{s,s'}\bra{+} [(\sum_{l_2> l_1}^{N_{\rm DD}}e^{-i\frac{(2l_1-1)\omega^{({\rm T})} \tau}{2}\hat{\sigma}_z^{({\rm T})}}\hat{M}_{s'}^{(1)}e^{-i\frac{2(l_2-l_1)\omega^{({\rm T})} \tau}{2}\hat{\sigma}_z^{({\rm T})}}\hat{M}_{s'}^{(1)}e^{-i\frac{(2(N_{\rm DD}-l_2)+1)\omega^{({\rm T})} \tau}{2}\hat{\sigma}_z^{({\rm T})}}\notag\\
&+\sum_{l=1}^{N_{\rm DD}}e^{-i\frac{(2l-1)\omega^{({\rm T})} \tau}{2}\hat{\sigma}_z^{({\rm T})}}\hat{M}_{s'}^{(2)}e^{-i\frac{(2(N_{\rm DD}-l)+1)\omega^{({\rm T})} \tau}{2}\hat{\sigma}_z^{({\rm T})}})^\dagger e^{-i(N_{\rm DD}\omega^{({\rm T})} \tau)\hat{\sigma}_z^{({\rm T})}}\notag\\
&+ e^{i(N_{\rm DD}\omega^{({\rm T})} \tau)\hat{\sigma}_z^{({\rm T})}}(\sum_{l_2> l_1}^{N_{\rm DD}}e^{-i\frac{(2l_1-1)\omega^{({\rm T})} \tau}{2}\hat{\sigma}_z^{({\rm T})}}\hat{M}_s^{(1)}e^{-i\frac{(2(l_2-l_1)+1)\omega^{({\rm T})} \tau}{2}\hat{\sigma}_z^{({\rm T})}}\hat{M}_s^{(1)}e^{-i\frac{(2(N_{\rm DD}-l_2)+1)\omega^{({\rm T})} \tau}{2}\hat{\sigma}_z^{({\rm T})}}\notag\\
&+\sum_{l=1}^{N_{\rm DD}}e^{-i\frac{(2l-1)\omega^{({\rm T})} \tau}{2}\hat{\sigma}_z^{({\rm T})}}\hat{M}_s^{(2)}e^{-i\frac{(2(N_{\rm DD}-l)+1)\omega^{({\rm T})} \tau}{2}\hat{\sigma}_z^{({\rm T})}})]\ket{+}\bra{s'}\hat{M}_x\ket{s}\\
=&L\sum_{j=1}^L\bra{+}e^{i(N_{\rm DD}\omega^{({\rm T})} \tau)\hat{\sigma}_z^{({\rm T})}} (-\sum_{l_2> l_1}^{N_{\rm DD}}e^{-i\frac{(2l_1-1)\omega^{({\rm T})} \tau}{2}\hat{\sigma}_z^{({\rm T})}}\left(\frac{2G}{\omega^{({\rm T})}}[1-\cos{(\omega^{({\rm T})}\tau)}]\times [B(\vec{r}_j),-A(\vec{r}_j),0]\cdot  \vec{\sigma}\right)\notag\\
&e^{-i\frac{2(l_2-l_1)\omega^{({\rm T})} \tau}{2}\hat{\sigma}_z^{({\rm T})}}\left(\frac{2G}{\omega^{({\rm T})}}[1-\cos{(\omega^{({\rm T})}\tau)}]\times [B(\vec{r}_j),-A(\vec{r}_j),0]\cdot  \vec{\sigma}\right)e^{-i\frac{(2(N_{\rm DD}-l_2)+1)\omega^{({\rm T})} \tau}{2}\hat{\sigma}_z^{({\rm T})}}\notag\\
&+\sum_{l=1}^{N_{\rm DD}}e^{-i\frac{(2l-1)\omega^{({\rm T})} \tau}{2}\hat{\sigma}_z^{({\rm T})}}\int_0^{\tau}d\lambda \int_0^\lambda d\xi  (-1)\times [\vec{n}_{j}(\lambda)\cdot  \vec{\sigma}][\vec{n}_{j}(\xi)\cdot  \vec{\sigma}]+[\vec{n}_{j}(-\xi)\cdot  \vec{\sigma}][\vec{n}_{j}(-\lambda)\cdot  \vec{\sigma}]\notag\\
&+\int_0^{\tau}d\lambda \int_0^{\tau} d\xi  [\vec{n}_{j}(\lambda)\cdot  \vec{\sigma}][\vec{n}_{j}(-\xi)\cdot  \vec{\sigma}]e^{-i\frac{(2(N_{\rm DD}-l)+1)\omega^{({\rm T})} \tau}{2}\hat{\sigma}_z^{({\rm T})}}) \ket{+}+c.c.\\
=&-2\times L\frac{4G^2}{[\omega^{({\rm T})}]^2}[1-\cos{(\omega^{({\rm T})}\tau)}]^2\sum_{l_2> l_1}^{N_{\rm DD}}\sum_{j=1}^L\bra{+}(\vec{n}_{l_1}\cdot  \vec{\sigma})(\vec{n}_{l_2}\cdot  \vec{\sigma})\ket{+}\notag\\
&+L \sum_{j=1}^L\sum_{l=1}^{N_{\rm DD}}\bra{+}e^{i\frac{(2(N_{\rm DD}-l)+1)\omega^{({\rm T})} \tau}{2}\hat{\sigma}_z^{({\rm T})}}[-\int_0^{\tau}d\lambda \int_0^\lambda d\xi  [[\vec{n}_{j}(\lambda)\cdot  \vec{\sigma}][\vec{n}_{j}(\xi)\cdot  \vec{\sigma}]+[\vec{n}_{j}(-\xi)\cdot  \vec{\sigma}][\vec{n}_{j}(-\lambda)\cdot  \vec{\sigma}]\notag\\
&+\int_0^{\tau}d\lambda \int_0^{\tau} d\xi  [\vec{n}_{j}(\lambda)\cdot  \vec{\sigma}][\vec{n}_{j}(-\xi)\cdot  \vec{\sigma}]e^{-i\frac{(2(N_{\rm DD}-l)+1)\omega^{({\rm T})} \tau}{2}\hat{\sigma}_z^{({\rm T})}}]+h.c. \ket{+}\\
=&- L\frac{4G^2}{[\omega^{({\rm T})}]^2}[1-\cos{(\omega^{({\rm T})}\tau)}]^2\sum_{ l_1\neq l_2}^{N_{\rm DD}}\sum_{j=1}^L[\vec{n}_{l_1}\cdot\vec{n}_{l_2}]\notag\\
&+L\frac{2G^2}{[\omega^{({\rm T})}]^2} \sum_{j=1}^L\sum_{l=1}^{N_{\rm DD}}\bra{+}e^{i\frac{(2(N_{\rm DD}-l)+1)\omega^{({\rm T})} \tau}{2}\hat{\sigma}_z^{({\rm T})}}[-[1-\cos{(\omega^{({\rm T})}\tau)}]^2[(A(\vec{r}_j))^2+(B(\vec{r}_j))^2]\mathbb{I}\notag\\
&-i[A(\vec{r}_j)\hat{\sigma}_x^{({\rm T})}+B(\vec{r}_j)\hat{\sigma}_y^{({\rm T})}]C(\vec{r}_j)[2\omega^{({\rm T})}\tau-2\sin{\omega^{({\rm T})}\tau}]\notag\\
&-i[(A(\vec{r}_j))^2+(B(\vec{r}_j))^2]\hat{\sigma}_z^{({\rm T})}[\omega^{({\rm T})}\tau-2\sin{\omega^{({\rm T})}\tau}+\sin{\omega^{({\rm T})}\tau}\cos{\omega^{({\rm T})}\tau}]e^{-i\frac{(2(N_{\rm DD}-l)+1)\omega^{({\rm T})} \tau}{2}\hat{\sigma}_z^{({\rm T})}}]+h.c. \ket{+}\\
=&- L\frac{4G^2}{[\omega^{({\rm T})}]^2}[1-\cos{(\omega^{({\rm T})}\tau)}]^2\sum_{ l_1,l_2}^{N_{\rm DD}}\sum_{j=1}^L[\vec{n}_{l_1}\cdot\vec{n}_{l_2}]\\
=&-L\frac{16G^2}{[\omega^{({\rm T})}]^2}[\sin{\frac{\omega^{({\rm T})}\tau}{2}}]^4\left[\frac{\sin{(N_{\rm DD}\omega^{({\rm T})}\tau)}}{\sin{(\omega^{({\rm T})}\tau)}}\right]^2\Gamma_{1,L}^{{\rm (sep)}}.
\end{align}
Therefore, we obtain
\begin{align}
&\frac{1}{2^{L}}\sum_{s,s'}\bra{+}\varepsilon^2 [(A_{s'}^{(1)})^\dagger A_s^{(1)}+(A_{s'}^{(2)})^\dagger e^{-i\frac{N_{\rm DD}\omega^{({\rm T})} t}{2}\hat{\sigma}_z^{({\rm T})}}+ e^{i\frac{N_{\rm DD}\omega^{({\rm T})} t}{2}\hat{\sigma}_z^{({\rm T})}}A_s^{(2)}]\ket{+}\bra{s'}\hat{M}_x\ket{s}\\
&=-\frac{32G^2}{[\omega^{({\rm T})}]^2}[\sin{\frac{\omega^{({\rm T})}\tau}{2}}]^4\left[\frac{\sin{(N_{\rm DD}\omega^{({\rm T})}\tau)}}{\sin{(\omega^{({\rm T})}\tau)}}\right]^2\Gamma_{1,L}^{{\rm (sep)}}.
\end{align}
If we prepare the initial state $\ket{\psi(0)}=\ket{-}\bigotimes _{j=1}^L \ket{+}_j$, we can confirm that the expectation value $\braket{\hat{M}_x}$ is the same.
Therefore, $\braket{\hat{M}_x}$ does not depend on whether the initial state is $\ket{+}$ or $\ket{-}$.

Note that the dephasing map $\mathcal{E}$ in Eq.~(\ref{eq:dephasingmap}) is Hermitian:
\begin{align}
\mathrm{Tr}[\hat{A}\mathcal{E}[\hat{\rho}]]&=\mathrm{Tr}[\mathcal{E}[\hat{A}]\hat{\rho}],\label{Hermitiannoise}
\end{align}
for any observable $\hat{A}$.
Therefore, we can obtain the expectation values and the variance with dephasing by using the results without dephasing:
\begin{align}
\mathcal{E}[\hat{M}_x]&=e^{-\left(\frac{t}{T_2^{\rm DD}}\right)^3}\times \hat{M}_x. 
\end{align}

The variance can be easily calculated in terms of the leading order:
\begin{align}
&\braket{(\hat{M}_x)^2}-\braket{\hat{M}_x}^2=L(1-e^{-2\left(\frac{t}{T_2^{\rm DD}}\right)^3})+O\left(\left[\frac{G}{\omega^{({\rm T})}}\right]^2\right).
\end{align}

\subsection{Derivation of the measurement probability in Eq.~(\ref{eq:measurementprobabilityGHZ})}
Let us calculate the measurement probability when the NV centers are prepared in a GHZ state.
As we explain in the main text,
our protocol is shown as below: In step 1, prepare the initial state $\ket{+}\otimes\ket{{\rm GHZ}}$.
In step 2, let the state evolve for a time $\tau$ with the effective total Hamiltonian $\hat{H}_{{\rm T}}+\hat{H}^{({\rm eff})}_{{\rm I}}$:\\
\begin{align}
\ket{\psi(\tau)}&=e^{-i(\hat{H}_{{\rm T}}+\hat{H}^{({\rm eff})}_{{\rm I}})\tau}\ket{+}\ket{{\rm GHZ}}\\
&=e^{-i(\frac{\omega^{({\rm T})} }{2}\hat{\sigma}_z^{({\rm T})}+\sum_{j=1}^L\hat{H}_j^{{\rm (T)}}\hat{\sigma}_{z,j}^{({\rm P})})\tau}\ket{+}\otimes\left(\frac{1}{\sqrt{2}}(\ket{00\cdots 0}+\ket{11\cdots 1}\right)\\
&=\frac{1}{\sqrt{2}}\left[e^{-i(\frac{\omega^{({\rm T})} }{2}\hat{\sigma}_z^{({\rm T})}+\sum_{j=1}^L\hat{H}_j^{{\rm (T)}})\tau}\ket{+}\otimes \ket{00\cdots 0}+e^{-i(\frac{\omega^{({\rm T})} }{2}\hat{\sigma}_z^{({\rm T})}-\sum_{j=1}^L\hat{H}_j^{{\rm (T)}})\tau}\ket{+}\otimes \ket{11\cdots 1}\right]\\
&=\frac{1}{\sqrt{2}}\left[\hat{U}_1\ket{+}\otimes \ket{00\cdots 0}+\hat{U}_2\ket{+}\otimes \ket{11\cdots 1}\right]
\end{align}
Subsequently,
perform a $\pi$ pulse on the probe spins:\\
\begin{align}
\ket{\psi'(\tau)}&=\left[\bigotimes_{j=1}^L\hat{\sigma}_{x,j}^{({\rm P})}\right]\ket{\psi(\tau)}\\
&=\frac{1}{\sqrt{2}}\left[\hat{U}_1\ket{+}\otimes \ket{11\cdots 1}+\hat{U}_2\ket{+}\otimes \ket{00\cdots 0}\right]
\end{align}
In step 3, again, let the state evolve for a time $\tau$ with $\hat{H}_{{\rm T}}+\hat{H}^{({\rm eff})}_{{\rm I}}$, and perform a $\pi$ pulse on the probe spins:\\
\begin{align}
&\ket{\psi'(2\tau)}=\left[\bigotimes_{j=1}^L\hat{\sigma}_{x,j}^{({\rm P})}\right]e^{-i(\hat{H}_{{\rm T}}+\hat{H}^{({\rm eff})}_{{\rm I}})\tau}\ket{\psi'(\tau)}\\
&=\frac{1}{\sqrt{2}}\left[\hat{U}_2\hat{U}_1\ket{+}\otimes \ket{00\cdots 0}+\hat{U}_1\hat{U}_2\ket{+}\otimes \ket{11\cdots 1}\right]
\end{align}
In step 4, measure the probe spins with the projection operator $\mathbb{I}_{{\rm T}}\otimes\ket{{\rm GHZ}}\bra{{\rm GHZ}}$. The probability $p({\rm GHZ})$ is given by
\begin{align}
p({\rm GHZ})&=\left\|\frac{1}{2}\left[\hat{U}_2\hat{U}_1+\hat{U}_1\hat{U}_2\right] \ket{+}\right\|^2\\
&=\frac{1}{2}+\frac{1}{2}\mathrm{Re}[\bra{+}\hat{U}^\dagger_2\hat{U}^\dagger_1\hat{U}_2\hat{U}_1\ket{+}].
\end{align}
We apply the second order perturbation to $\hat{U}_2\hat{U}_1$ and $\hat{U}_2^\dagger \hat{U}_1^\dagger$ of Eq. (\ref{perturbationformula})
for $A=\pm\frac{\omega^{({\rm T})} }{2}\hat{\sigma}_z^{({\rm T})}$ and $\varepsilon B=\pm\sum_{j=1}^L\hat{H}_j^{{\rm (T)}}$:
\begin{align}
\hat{U}_2\hat{U}_1&=e^{-i(\frac{\omega^{({\rm T})} }{2}\hat{\sigma}_z^{({\rm T})}-\sum_{j=1}^L\hat{H}_j^{{\rm (T)}})\tau}e^{-i(\frac{\omega^{({\rm T})} }{2}\hat{\sigma}_z^{({\rm T})}+\sum_{j=1}^L\hat{H}_j^{{\rm (T)}})\tau}\\
&=e^{-i\frac{\omega^{({\rm T})} \tau}{2}\hat{\sigma}_z^{({\rm T})}}(\mathbb{I}_{{\rm T}}+\varepsilon \hat{U}^{(1)}_a+\varepsilon^2 \hat{U}^{(2)}_a)e^{-i\frac{\omega^{({\rm T})} \tau}{2}\hat{\sigma}_z^{({\rm T})}},
\end{align}
where we define
\begin{align}
\varepsilon F_a(\lambda)&=e^{i\frac{\omega^{({\rm T})} }{2}\hat{\sigma}_z^{({\rm T})}\lambda }\sum_{j=1}^L\hat{H}_j^{{\rm (T)}}e^{-i\frac{\omega^{({\rm T})} }{2}\hat{\sigma}_z^{({\rm T})}\lambda}\\
&=G\sum_{j=1}^L[A(\vec{r}_j)(\hat{\sigma}_x^{({\rm T})}\cos{(\omega^{({\rm T})}\lambda)}- \hat{\sigma}_y^{({\rm T})}\sin{(\omega^{({\rm T})}\lambda)})+B(\vec{r}_j)(\hat{\sigma}_y^{({\rm T})}\cos{(\omega^{({\rm T})}\lambda)}+ \hat{\sigma}_x^{({\rm T})}\sin{(\omega^{({\rm T})}\lambda)}) +C(\vec{r}_j)\hat{\sigma}_z^{({\rm T})} ]\\
&=\sum_{j=1}^L \vec{n}_j(\lambda) \cdot \vec{\sigma}\\
\varepsilon \hat{U}^{(1)}_a=&i\varepsilon \int_0^{\tau}d\lambda [F_a(\lambda)-F_a(-\lambda)]
=i\int_0^{\tau} d\lambda \sum_{j=1}^L[\vec{n}_j(\lambda) -\vec{n}_j(-\lambda)]\cdot  \vec{\sigma}\\
\varepsilon^2\hat{U}^{(2)}_a=&-\varepsilon^2 \int_0^{\tau}d\lambda \int_0^\lambda d\xi  [F_a(\lambda)F_a(\xi)+F_a(-\xi)F_a(-\lambda)]+\varepsilon^2 \int_0^{\tau}d\lambda \int_0^{t/2} d\xi  F_a(\lambda)F_a(-\xi),
\end{align}
and
\begin{align}
\hat{U}_2^\dagger \hat{U}_1^\dagger&=e^{i(\frac{\omega^{({\rm T})} }{2}\hat{\sigma}_z^{({\rm T})}-\sum_{j=1}^L\hat{H}_j^{{\rm (T)}})\tau}e^{i(\frac{\omega^{({\rm T})} }{2}\hat{\sigma}_z^{({\rm T})}+\sum_{j=1}^L\hat{H}_j^{{\rm (T)}})\tau}\\
&=e^{-i(-\frac{\omega^{({\rm T})} }{2}\hat{\sigma}_z^{({\rm T})}+\sum_{j=1}^L\hat{H}_j^{{\rm (T)}})\tau}e^{-i(-\frac{\omega^{({\rm T})} }{2}\hat{\sigma}_z^{({\rm T})}-\sum_{j=1}^L\hat{H}_j^{{\rm (T)}})\tau}\\
&=e^{i\frac{\omega^{({\rm T})} \tau}{2}\hat{\sigma}_z^{({\rm T})}}(\mathbb{I}_{{\rm T}}+\varepsilon \hat{U}^{(1)}_b+\varepsilon^2 \hat{U}^{(2)}_b)e^{i\frac{\omega^{({\rm T})} \tau}{2}\hat{\sigma}_z^{({\rm T})}},
\end{align}
where we define
\begin{align}
\varepsilon F_b(\lambda)&=-\varepsilon F_a(-\lambda)=-\sum_{j=1}^L \vec{n}_j(-\lambda) \cdot \vec{\sigma}\\
\varepsilon \hat{U}^{(1)}_b=&i\varepsilon \int_0^{\tau}d\lambda [F_b(\lambda)-F_b(-\lambda)]
=\varepsilon \hat{U}^{(1)}_a\\
\varepsilon^2\hat{U}^{(2)}_b=&-\varepsilon^2 \int_0^{\tau}d\lambda \int_0^\lambda d\xi  [F_b(\lambda)F_b(\xi)+F_b(-\xi)F_b(-\lambda)]+\varepsilon^2 \int_0^{\tau}d\lambda \int_0^{\tau} d\xi  F_b(\lambda)F_b(-\xi)
\end{align}
and finally we will calculate
\begin{align}
\hat{U}_2^\dagger \hat{U}_1^\dagger\hat{U}_2\hat{U}_1
&=e^{i\frac{\omega^{({\rm T})} \tau}{2}\hat{\sigma}_z^{({\rm T})}}(\mathbb{I}_{{\rm T}}+\varepsilon [\hat{U}^{(1)}_a+\hat{U}^{(1)}_b]+\varepsilon^2[\hat{U}^{(1)}_b\hat{U}^{(1)}_a+ \hat{U}^{(2)}_a+\hat{U}^{(2)}_b])e^{-i\frac{\omega^{({\rm T})} \tau}{2}\hat{\sigma}_z^{({\rm T})}}.
\end{align}
and therefore we obtain
\begin{align}
\varepsilon[\hat{U}^{(1)}_a+ \hat{U}^{(1)}_b]
&=2i \varepsilon \int_0^{\tau}d\lambda [F(\lambda)-F(-\lambda)]\\
&=2i \sum_{j=1}^L\int_0^{\tau}d\lambda G[-A(\vec{r}_j) \hat{\sigma}_y^{({\rm T})}\sin{(\omega^{({\rm T})}\lambda)}+B(\vec{r}_j) \hat{\sigma}_x^{({\rm T})}\sin{(\omega^{({\rm T})}\lambda)}  ]\\
&=2iG\frac{1-\cos{(\omega^{({\rm T})}\tau)}}{\omega^{({\rm T})}}\sum_{j=1}^L[-A(\vec{r}_j) \hat{\sigma}_y^{({\rm T})}+B(\vec{r}_j) \hat{\sigma}_x^{({\rm T})}]
\end{align}
Moreover, we calculate each of the terms $\varepsilon^2[\hat{U}^{(1)}_b\hat{U}^{(1)}_a+ \hat{U}^{(2)}_a+\hat{U}^{(2)}_b]$:
\begin{align}
&\varepsilon^2 \hat{U}^{(1)}_b\hat{U}^{(1)}_a=(\varepsilon \hat{U}^{(1)}_a)^2\\
&=\left[i \varepsilon \int_0^{\tau}d\lambda [F(\lambda)-F(-\lambda)]\right]^2\\
&=-\left[2\sum_{j=1}^L\int_0^{\tau}d\lambda G[-A(\vec{r}_j) \hat{\sigma}_y^{({\rm T})}\sin{(\omega^{({\rm T})}\lambda)}+B(\vec{r}_j) \hat{\sigma}_x^{({\rm T})}\sin{(\omega^{({\rm T})}\lambda)}  ]\right]^2\\
&=-4G^2\frac{[1-\cos{(\omega^{({\rm T})}\tau)}]^2}{[\omega^{({\rm T})}]^2}\left[\sum_{j=1}^L[-A(\vec{r}_j) \hat{\sigma}_y^{({\rm T})}+B(\vec{r}_j) \hat{\sigma}_x^{({\rm T})} ]\right]^2\\
&=-4G^2\frac{[1-\cos{(\omega^{({\rm T})}\tau)}]^2}{[\omega^{({\rm T})}]^2}\sum_{j_1,j_2=1}^L[A(\vec{r}_{j_1})A(\vec{r}_{j_2})+B(\vec{r}_{j_1})B(\vec{r}_{j_2})]\mathbb{I}_{{\rm T}}+i[A(\vec{r}_{j_1})B(\vec{r}_{j_2})-A(\vec{r}_{j_2})B(\vec{r}_{j_1})]\hat{\sigma}_z^{({\rm T})}\\
&=-4G^2\frac{[1-\cos{(\omega^{({\rm T})}\tau)}]^2}{[\omega^{({\rm T})}]^2}[\bar{A}^2+\bar{B}^2]\mathbb{I}_{{\rm T}}\\
&=-\frac{16G^2}{[\omega^{({\rm T})}]^2}[\sin{\frac{\omega^{({\rm T})}\tau}{2}}]^4\Gamma_{1,L}^{{\rm (ent)}}\mathbb{I}_{{\rm T}},
\end{align}
where we define
\begin{align}
\bar{A}&= \sum_{j=1}^LA(\vec{r}_{j}),\quad
\bar{B}=\sum_{j=1}^LB(\vec{r}_{j})\\
\bar{A}^2+\bar{B}^2&=\Gamma_{1,L}^{{\rm (ent)}}.
\end{align}
Also, we calculate the other terms:
\begin{align}
\varepsilon^2 [\hat{U}^{(2)}_a+\hat{U}^{(2)}_b]
&=-\varepsilon^2\left[\int_0^{\tau}d\lambda \int_0^\lambda d\xi  [F(\lambda)F(\xi)+F(-\xi)F(-\lambda)] -\int_0^{\tau}d\lambda \int_0^{\tau} d\xi  F(\lambda)F(-\xi) \right]\notag\\
&-\varepsilon^2\left[\int_0^{\tau}d\lambda \int_0^\lambda d\xi  [F(-\lambda)F(-\xi)+F(\xi)F(\lambda)] -\int_0^{\tau}d\lambda \int_0^{\tau} d\xi  F(-\lambda)F(\xi) \right]\\
&=-\varepsilon^2\int_0^{\tau}d\lambda \int_0^{\tau} d\xi [F(\lambda)F(\xi)+F(-\lambda)F(-\xi)-F(\lambda)F(-\xi)-F(-\lambda)F(\xi)]\\
&=-\left[\varepsilon\int_0^{\tau}d\lambda [F(\lambda)-F(-\lambda)]\right]^2 \\
&=\varepsilon^2 \hat{U}^{(1)}_b\hat{U}^{(1)}_a\\
&=-\frac{16G^2}{[\omega^{({\rm T})}]^2}[\sin{\frac{\omega^{({\rm T})}\tau}{2}}]^4\Gamma_{1,L}^{{\rm (ent)}}\mathbb{I}_{{\rm T}}. 
\end{align}
Therefore, we obtain
\begin{align}
\hat{U}_2^\dagger \hat{U}_1^\dagger\hat{U}_2\hat{U}_1
&=e^{i\frac{\omega^{({\rm T})} \tau}{2}\hat{\sigma}_z^{({\rm T})}}\left(\mathbb{I}_{{\rm T}}+2iG\frac{1-\cos{(\omega^{({\rm T})}\tau)}}{\omega^{({\rm T})}}[\bar{B} \hat{\sigma}_x^{({\rm T})}-\bar{A} \hat{\sigma}_y^{({\rm T})}]-\frac{32G^2}{[\omega^{({\rm T})}]^2}[\sin{\frac{\omega^{({\rm T})}\tau}{2}}]^4\Gamma_{1,L}^{{\rm (ent)}}\mathbb{I}_{{\rm T}}\right)e^{-i\frac{\omega^{({\rm T})} \tau}{2}\hat{\sigma}_z^{({\rm T})}}.
\end{align}
and
\begin{align}
p({\rm GHZ})
&= 1-\frac{16G^2}{[\omega^{({\rm T})}]^2}[\sin{\frac{\omega^{({\rm T})}\tau}{2}}]^4\Gamma_{1,L}^{{\rm (ent)}}+O\left(\left[\frac{G}{\omega^{({\rm T})}}\right]^3\right).
\end{align}
If we prepare the initial state  $\ket{\psi(0)}=\ket{-}\otimes\ket{{\rm GHZ}}$, we can confirm that the measurement probability $p({\rm GHZ})$ is the same.
Therefore, $p({\rm GHZ})$ does not depend on whether the initial state is $\ket{+}$ or $\ket{-}$.

Note that the dephasing map $\mathcal{E}$ in Eq.~(\ref{eq:dephasingmap}) is Hermitian as we explained in Eq.~(\ref{Hermitiannoise}).
Therefore, we can obtain the measurement probability with dephasing by using the results without dephasing and measuring the following observable:
\begin{align}
\mathcal{E}[\ket{{\rm GHZ}}\bra{{\rm GHZ}}]&=\frac{1}{2}\left[\ket{0\cdots 0}\bra{0\cdots 0}+\ket{1\cdots 1}\bra{1\cdots 1}+e^{-L\left(\frac{t}{T_2^{\rm echo}}\right)^3}[\ket{0\cdots 0}\bra{1\cdots 1}+\ket{1\cdots 1}\bra{0\cdots 0}]\right]. 
\end{align}
Hence,
\begin{align}
p({\rm GHZ})&=
\mathrm{Tr}\Bigl[\mathcal{E}[\ket{\psi'(2\tau)}\bra{\psi'(2\tau)}]\mathbb{I}_{{\rm T}}\otimes\ket{{\rm GHZ}}\bra{{\rm GHZ}}\Bigr]\\
&=\mathrm{Tr}\Bigl[\ket{\psi'(2\tau)}\bra{\psi'(2\tau)}\mathbb{I}_{{\rm T}}\otimes\mathcal{E}[\ket{{\rm GHZ}}\bra{{\rm GHZ}}]\Bigr]\\
&=\frac{1}{2}+\frac{e^{-L\left(\frac{2\tau}{T_2^{\rm echo}}\right)^3}}{2}\mathrm{Re}[\bra{+}\hat{U}^\dagger_2\hat{U}^\dagger_1\hat{U}_2\hat{U}_1\ket{+}]. 
\end{align}

\section{Explicit form of geometric factor}\label{sectionapp:Explicit}
For the case of the separable states, after the calculation of the integral in Eq.~(\ref{eq:separablegeometric}), we obtain
\begin{align}F_{1,L}^{{\rm (sep)}}(\tilde{r}_{{\rm max}},\tilde{z}_{{\rm max}})&= \frac{1}{3}-\frac{1}{3\tilde{z}_{\rm max}^3}+\frac{-5\tilde{r}_{{\rm max}}^5\tilde{z}_{\rm max}+8\tilde{r}_{{\rm max}}^3\tilde{z}_{\rm max}^3+5\tilde{r}_{{\rm max}}\tilde{z}_{\rm max}^5+5(\tilde{r}_{{\rm max}}^2+\tilde{z}_{\rm max}^2)^3{\rm ArcTan}[\tilde{z}_{\rm max}/\tilde{r}_{{\rm max}}]}{16\tilde{r}_{{\rm max}}^3(\tilde{r}_{{\rm max}}^2+\tilde{z}_{\rm max}^2)^3} \notag\\
&-\frac{-5\tilde{r}_{{\rm max}}^5+8\tilde{r}_{{\rm max}}^3+5\tilde{r}_{{\rm max}}+5(\tilde{r}_{{\rm max}}^2+1)^3{\rm ArcTan}[1/\tilde{r}_{{\rm max}}]}{16\tilde{r}_{{\rm max}}^3(\tilde{r}_{{\rm max}}^2+1)^3},
\end{align}
where we define dimensionless variables: 
$
\tilde{r}_{{\rm max}}=\frac{r_{{\rm max}}}{z_{{\rm min}}},
\tilde{z}_{{\rm max}}=\frac{z_{{\rm max}}}{z_{{\rm min}}}
$.

For the case of GHZ states, the geometric factor can be simplified as
\begin{align}
 F_{1,L}^{{\rm (ent)}}(\tilde{r}_{{\rm max}},\tilde{z}_{{\rm max}})
&=\left[-\log{\frac{\sqrt{\tilde{r}_{{\rm max}}^2+\tilde{z}_{\rm max}^2}+\tilde{r}_{{\rm max}}}{\sqrt{\tilde{r}_{{\rm max}}^2+\tilde{z}_{\rm max}^2}-\tilde{r}_{{\rm max}}}}\log{\frac{\sqrt{\tilde{r}_{{\rm max}}^2+1}+\tilde{r}_{{\rm max}}}{\sqrt{\tilde{r}_{{\rm max}}^2+1}-\tilde{r}_{{\rm max}}}}+\left[\frac{2\tilde{r}_{{\rm max}}}{(\tilde{r}_{{\rm max}}^2+\tilde{z}_{\rm max}^2)^{1/2}}-\frac{2\tilde{r}_{{\rm max}}}{(\tilde{r}_{{\rm max}}^2+1)^{1/2}} \right]\right]^2.
\end{align}

\end{widetext}

\end{document}